\documentclass[structabstract]{aa}  
\usepackage[modulo,switch]{lineno}
\usepackage{graphicx}
\usepackage{natbib}
\usepackage{longtable}
\usepackage{txfonts}
\begin{document}
%
   \title{Oscillation mode lifetimes of red giants observed during the initial and first anticentre long run of CoRoT\thanks{The CoRoT space mission which was developed and is operated by the French space agency CNES, with participation of ESA's RSSD and Science Programmes, Austria, Belgium, Brazil, Germany, and Spain. Light curves can be retrieved from the CoRoT archive: http://idoc-corot.ias.u-psud.fr.}}

   \author{S. Hekker\inst{1}\fnmsep\inst{2} \and C. Barban\inst{3} \and F. Baudin\inst{4} \and J. De Ridder\inst{2} \and T. Kallinger\inst{5} \and T. Morel\inst{6} \and W.~J. Chaplin\inst{1} \and Y. Elsworth\inst{1}}

   \institute{University of Birmingham, School of Physics and Astronomy, Edgbaston, Birmingham B15 2TT, United Kingdom\\
              \email{saskia@bison.ph.bham.ac.uk}
         \and Instituut voor Sterrenkunde, K.U. Leuven, Celestijnenlaan 200D, 3001 Leuven, Belgium
         \and LESIA, UMR8109, Universit\'e Pierre et Marie Curie, Universit\'e Denis Diderot, Observatoire de Paris, 92195 Meudon Cedex, France
         \and Institute d'Astrophysique Spatiale, UMR 8617, Universit\'e Paris XI, B\^atiment 121, 91405 Orsay Cedex, France
         \and Department of Physics and Astronomy, University of British Colombia, 6224 Agricultural Road, Vancouver, BC V6T 1Z1, Canada
         \and Institut d'Astrophysique et de G\'eophysique, Universit\'e de Li\`ege, All\'ee du 6 Ao\^ut, 4000 Li\`ege, Belgium
}

   \date{Received ; accepted}

 
  \abstract
   {Long timeseries of data increase the frequency resolution in the power spectrum. This allows for resolving stochastically excited modes with long mode lifetimes, as well as features that are close together in frequency. The CoRoT fields observed during the initial run and second long run partly overlap, and stars in this overlapping field observed in both runs are used to create timeseries with a longer timespan than available from the individual runs.}
   {We aim to measure the mode lifetimes of red giants and compare them with theoretical predictions. We also investigate the dependence of the mode lifetimes on frequency and the degree of the oscillation modes.}
   {We perform simulations to investigate the influence of the gap in the data between the initial and second long run, the total length of the run and the signal-to-noise ratio on the measured mode lifetime.  This provides us with a correction factor to apply to the mode lifetimes measured from a maximum likelihood fit to the oscillation frequencies.}
   {We find that the length of the timeseries, the signal-to-noise ratio and possible gaps do have a non-negligible effect on the measurements of the mode lifetime of stochastically excited oscillation modes, but that we can correct for it. For the four stars for which we can perform a fit of the oscillation frequencies, we find that the mode lifetimes depend on frequency and on degree of the mode, which is in quantitative agreement with theoretical predictions.}
   {}


   \maketitle
%

\section{Introduction}
The uninterupted long timeseries of high-precision photometric data obtained with the CoRoT satellite \citep{baglin2006a} opened many new possibilities for red-giant asteroseismology. Red giants have extended turbulent atmospheres in which solar-like oscillations are stochastically excited and intrinsically damped. These oscillations have periods of a few hours, and can have mode lifetimes ranging from days to hundreds of days \citep{dupret2009}. The mode lifetime of a stochastically excited mode is the characteristic time it takes for an excited mode to damp exponentially. This is an important parameter as it is directly related to the mode inertia and therefore the cumulated work integral. From this work integral we can derive in which part of the star the oscillations are driven and damped \citep[see][for more details]{dupret2009}. \citet{dupret2009} also predict that mode lifetimes depend on the degree of the mode and on frequency. This dependence follows similar trends for stars in different evolutionary phases, although the exact variation of the mode lifetimes with frequency changes depending on the evolutionary phase of the star. This frequency and mode dependence of the mode lifetime has not yet been observed in red giants and will be investigated here.

The mode lifetimes ($\tau$, e-folding time for the amplitude) can be measured from the width (FWHM) of the frequency peaks ($\Gamma$) of an oscillation mode in the power spectrum as
\begin{equation}
\tau = \frac{1}{\pi\Gamma}.
\label{lifetime}
\end{equation}
To measure the mode lifetime it is therefore important to have sufficient frequency resolution to resolve the linewidth of the oscillation mode. 
Examples of stars with long ($>$~50~day) and short ($<$~20~day) lifetimes were first shown by \citet{deridder2009} and \citet{carrier2010} and investigated for a larger number of stars by \citet{baudin2010}. Where \citet{baudin2010} propose an automated procedure to fit the line width and height for the three most prominent oscillation modes in a star, with the purpose of a statistical study of the amplitudes of the modes, we adopt here an individual analysis for a smaller sample of stars, but with longer timeseries. The data we have at our disposal are described in Sect. 2. 
Because we do not have `infinitely' long timeseries and there are considerable gaps in those series, we perform simulations to identify a possible bias on $\Gamma$ in Eq.~\ref{lifetime}.  This bias will be taken into account in computing the `unbiased' mode lifetime. These simulations are described in Sect. 3.

We performed a detailed analysis of four stars. For each star we then computed an \'{e}chelle diagram \citep[frequency versus frequency modulo large frequency separation,][]{grec1983}. From this \'{e}chelle diagram we aim to identify the degree of the oscillation modes, or at least identify modes with the same degree to investigate the dependence of the mode lifetime on degree. This is possible when the observed oscillation modes have high-order ($n$) and low-degree ($\ell$) and follow the asymptotic relation \citet{tassoul1980}, for which we use here the following equation:
\begin{equation}
\nu_{n,\ell} \approx \Delta \nu \left (n+\frac{1}{2}\ell+\epsilon \right )-\ell(\ell+1)D_0,
\label{asymptot}
\end{equation}
with $\nu$ the oscillation frequency, $\Delta \nu$ the large separation between modes of same degree and consecutive orders, which is inversely proportional to the sound travel time through the star. $D_0$ is sensitive to deeper layers in the star and $\epsilon$ to the surface layers. The large separation can be estimated from the relation between the frequency of maximum oscillation power ($\nu_{\rm max}$) and $\Delta \nu$ as found in an ensemble study of the first long run of CoRoT observations \citep{hekker2009}:
\begin{equation}
\Delta \nu=\Delta \nu_{\odot} \left (\frac{\nu_{\rm max}}{\nu_{\rm max \odot}} \right )^{0.78},
\label{dnunumax}
\end{equation}
or similar relations by \citet{stello2009} and \citet{mosser2010}. We use this relation to check whether we indeed found $\Delta \nu$ and not a (sub)multiple of $\Delta \nu$.

In general three small separations are defined: $\delta \nu_{02}$ is the spacing between adjacent modes with $\ell$~=~0 and $\ell$~=~2, $\delta \nu_{13}$ is the spacing between adjacent modes with $\ell$~=~1 and $\ell$~=~3, and $\delta \nu_{01}$ is the amount by which  the $\ell$~=~1 modes are offset from the midpoints between the $\ell$~=~0 mode on either side. If Eq.~\ref{asymptot} holds then $\delta \nu_{02}$ = 6$D_0$, $\delta \nu_{13}$ = 10$D_0$ and $\delta \nu_{01}$ = 2$D_0$, although there is evidence that for the majority of red-giant stars $\delta \nu_{01}$ is negative \citep{carrier2010} and \citet{bedding2010}, i.e. the $\ell$~=~1 modes are located on the right side of the midpoints between the $\ell$~=~0 modes instead of on the left side of the midpoint as is the case for the Sun. This is most likely caused by stellar evolution and the asymptotic relation given above may not be valid for evolved stars. 

\section{Observations}

\begin{figure}
\begin{minipage}{\linewidth}
\centering
\includegraphics[width=\linewidth]{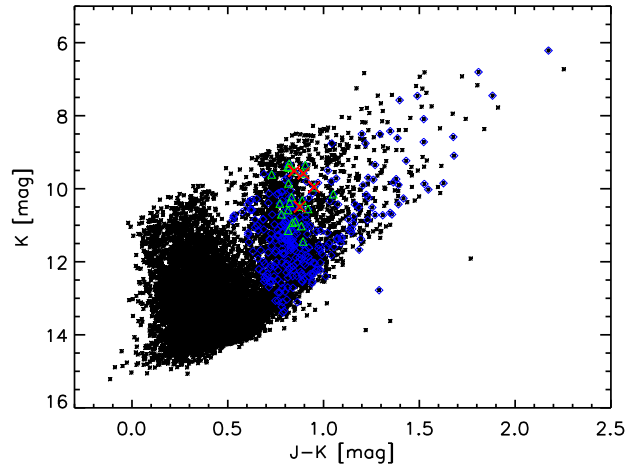}
\end{minipage}
\caption{Colour-magnitude diagram of the field stars observed during LRa01 using the 2MASS $J$ and $Ks$ photometric passbands (black dots) \citep{deleuil2009}. The blue diamonds represent the red-giant stars also observed during IR, the green triangles represent the red-giant candidates which show clear power excess and the red crosses represent the four stars for which we could fit the individual oscillation modes.}
\label{colmag}
\end{figure}

\begin{figure}
\begin{minipage}{\linewidth}
\centering
\includegraphics[width=\linewidth]{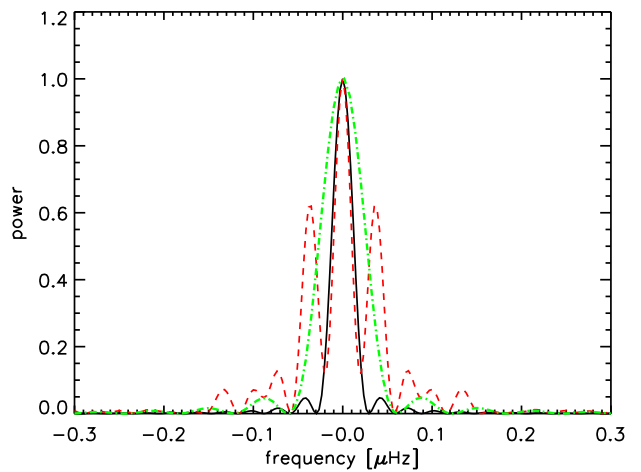}
\end{minipage}
\caption{Window functions for a 394 day timeseries (black solid line), the same timeseries with a large gap, i.e., no data between day 58 and day 263 (red dashed line) and for the same timeseries with the data of the first 58 and last 131 days `merged' (green dot-dashed line). The window functions are oversampled by a factor of ten for visual purposes.}
\label{windowfunc}
\end{figure}

For the present analyses we use CoRoT data taken during the Initial Run (IR) and first long run in the anti-centre direction (LRa01) in the exo-planet field. For information on CoRoT and technical details about the CoRoT data reduction we refer to \citet{baglin2006}. The IR lasted 58 days from 2$^{\rm nd}$ February to 31$^{\rm st}$ March 2007 during which the satellite was pointed towards a field with the centre at equatorial coordinates ($\alpha$, $\delta$) = (102.60$^{\circ}$, $-$1.7$^{\circ}$). During the 131 day long LRa01, which lasted from October 2007 to March 2008, the satellite observed a field around equatorial coordinates ($\alpha$, $\delta$) = (102.72$^{\circ}$, $-$0.2$^{\circ}$). The typical timestep of the observations is 32 s, although some targets are observed at 512 s intervals.

Visual and near-IR photometry are available for all stars in the IR and LRa01 field \citep{deleuil2009}. These colours are affected by reddening, but as shown by \citet{bessell1988} near-IR colours are least affected and provide a first estimate of the spectral type. A $J - K$ versus $K$ colour magnitude diagram is shown in Fig.~\ref{colmag}.

We use these colours to distinguish between main-sequence and red-giant stars. In this way we found 305 red-giant candidates based on colours (blue diamonds in Fig.~\ref{colmag}), which are observed during both the IR and LRa01. We computed a combined timeseries for these candidates, in which we `merged' the timeseries. Simulations to investigate the effect of merging the IR and LRa01 datasets are described in Sect. 3. Then we computed power spectra and global fits to them. These fits are composed of a power law to account for the background of the spectra due to effects such as granulation, and a Gaussian fit to the oscillation power excess, similar to the fitting described by \citet{hekker2009}. For 24 stars (green triangles in Fig.~\ref{colmag}) we found clear power excess and we investigated the oscillations in more detail. For four stars the noise level was low enough to perform  fitting of individual oscillation modes. These four stars are indicated with red crosses in the colour-magnitude diagram in Fig.~\ref{colmag} and the detailed analysis of these stars is described in Sect. 4.

\begin{figure}
\begin{minipage}{\linewidth}
\centering
\includegraphics[width=\linewidth]{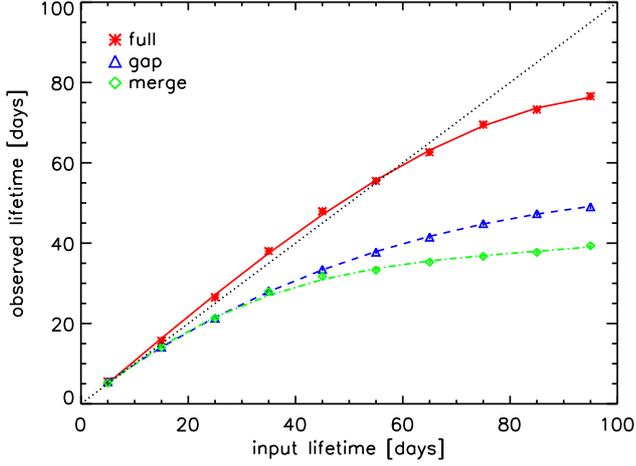}
\end{minipage}
\caption{Average of the lifetimes observed for 1000 simulated realisations as a function of their input lifetime. The `full', `gap' and `merge' cases are indicated with red asterisks, blue triangles and green diamonds respectively. Errors on the mean are plotted, but are in most cases smaller than the symbol size. The black dotted line is the one-to-one relation.}
\label{lifeavgobs}
\end{figure}

\begin{figure}
\begin{minipage}{\linewidth}
\centering
\includegraphics[width=\linewidth]{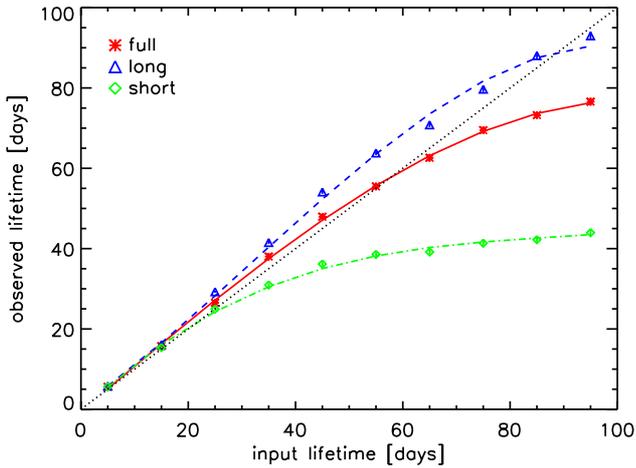}
\end{minipage}
\caption{Same as Fig.~\ref{lifeavgobs} but here the red asterisks, blue triangles and green diamonds indicate the `full', `long' and `short' cases, respectively.}
\label{lifeavgtest1}
\end{figure}

\section{Simulations}
The large gap of 205 days between the initial run (58 days) and the first long run in the anti-centre direction (131 days) has a large influence on the window function (Fig.~\ref{windowfunc}) by introducing prominent sidelobes, and these have to be taken into account in the analysis of the data. To investigate how we should treat the gap and what effect the gap has on the resulting parameters obtained from a fit to the power spectrum we performed simulations using the simulator described by \citet{chaplin1997}. First we performed simulations of noise-free timeseries which cover the full timespan of 394 days with 512 second cadence. Each timeseries has one stochastically excited oscillation mode at a frequency in the frequency range of red giants observed with CoRoT, i.e., 5, 15, 25, ...., 95 $\mu$Hz and for each frequency we simulated ten different mode lifetimes of 5 up to 95 days. For each combination of frequency and mode lifetime we simulated 1000 realisations.

In these simulations of the full timespan we introduced a gap by removing all data between day 58 and day 263, but not altering any of the time stamps. In addition, we also merge the timeseries data of the first 58 days and last 131 days by removing the gap and altering the timestamps (see Fig.~\ref{windowfunc} for the window functions of these scenarios). The latter is an option for red giants as these have stochastically excited modes with finite mode lifetimes. These stochastic modes do not have a coherent phase and one may therefore in principle merge data sets with different epochs.

We analysed for all three cases, i.e., `full', `gap' and `merge', the individual realisations. In all instances we fitted the power spectrum using a maximum likelihood technique \citep[e.g.,][]{anderson1990}, where the final model $M$ used for comparison is a convolution of the model with the power spectrum of the observed window function of the data set, normalised to unit total area (see Eq.~\ref{model}). This takes the redistribution of power caused by gaps in the data into account. 
\begin{equation}
M=\left(\sum_j \frac{A_j}{1+((\nu_{\rm cen,\it j} -\nu)/B_j)^2}\right) \ast \rm window
\label{model}
\end{equation}
In this model, each of the $j$ oscillation peaks are fitted with the amplitude ($A$), the central frequency ($\nu_{\rm cen}$) and FWHM ($B$) as free parameters. 
To maximise the possibility of obtaining the best values, i.e., the solution of the global instead of local maximum, we use different input parameters for the linewidth and height of the frequency peak. From the total of twelve fits performed for each timeseries we choose the one with the maximum likelihood. Before we used the result we also tested if we could consider the fitted function as a good approximation of the parent distribution by means of reduced figure-of-merit function \citep[MERIT$_{\rm red}$, see][and references therein]{anderson1990}:
\begin{equation}
{\rm MERIT_{red}} = \frac{1}{f} \sum_i \left (\frac{O_i-M_i}{M_i} \right)^2,
\label{merit}
\end{equation}
with $O_i$ the observed power at frequency $i$ and $M_i$ the model at the same frequency and $f$ the number of degrees of freedom, i.e., the difference between the number of independent data points in the fit and the number of free parameters in the model. A fit with MERIT$_{\rm red}$ less than one can be considered a good fit to the data.

\begin{figure}
\begin{minipage}{\linewidth}
\centering
\includegraphics[width=\linewidth]{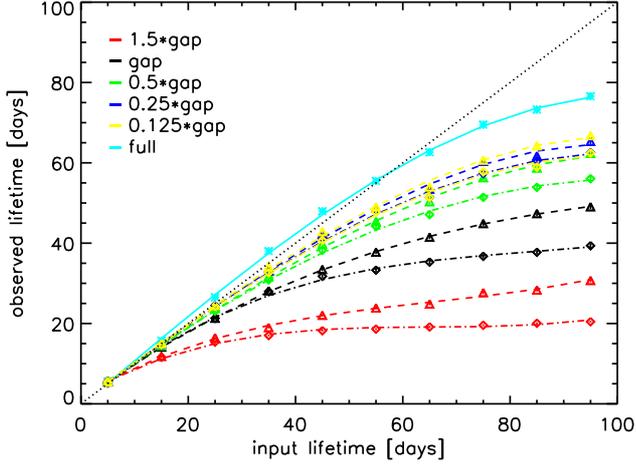}
\end{minipage}
\caption{Same as Fig.~\ref{lifeavgobs} for the `gap' (triangles) and `merge' (diamonds) cases with 1.5, 1, 0.5, 0.25 and 0.125 times the gap in the observations. These are indicated in red, black, green, blue and yellow, respectively. The results for the full timeseries are indicated with the light blue asterisks.}
\label{lifeavgtest2}
\end{figure}

\begin{figure}
\begin{minipage}{\linewidth}
\centering
\includegraphics[width=\linewidth]{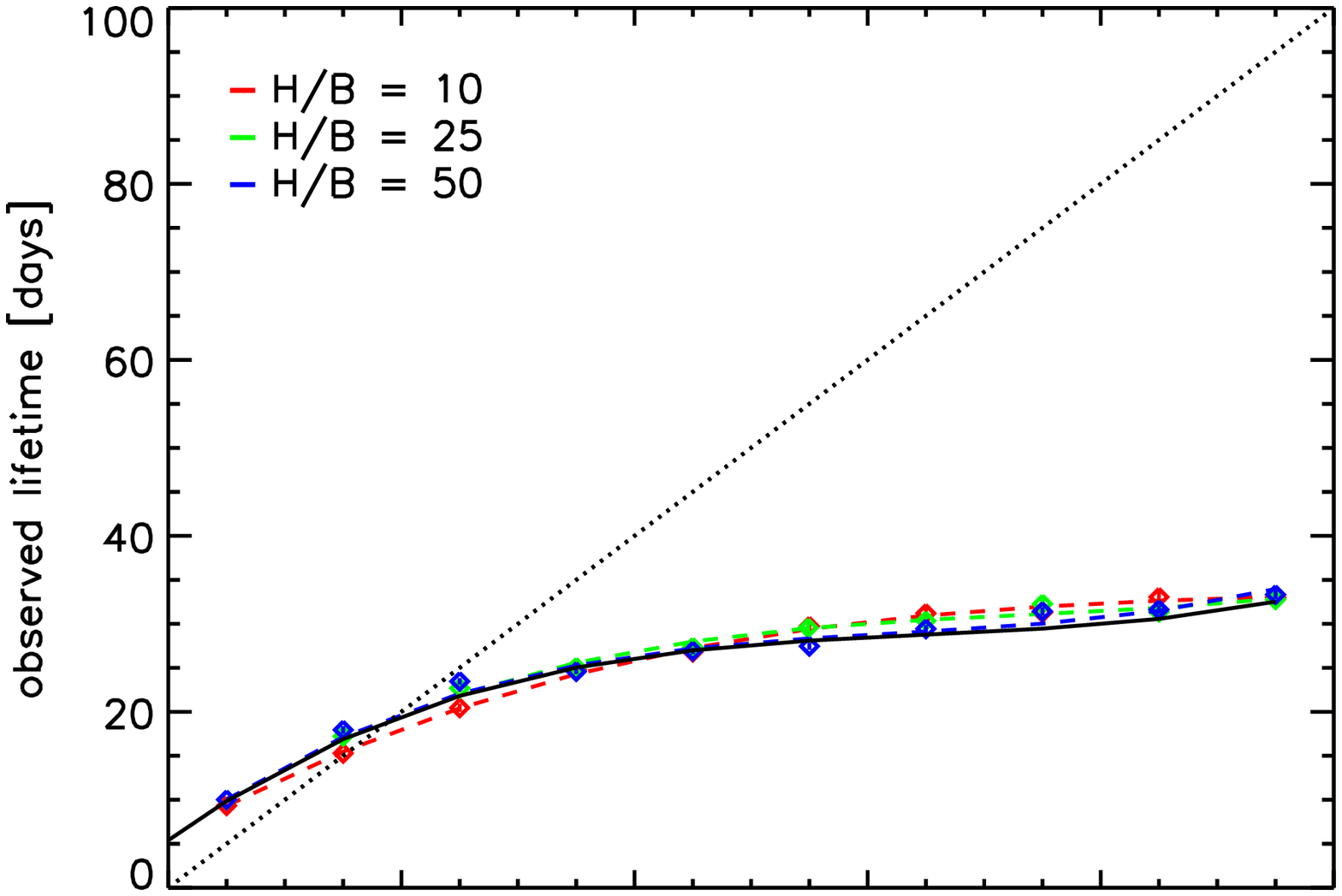}
\end{minipage}
\begin{minipage}{\linewidth}
\centering
\includegraphics[width=\linewidth]{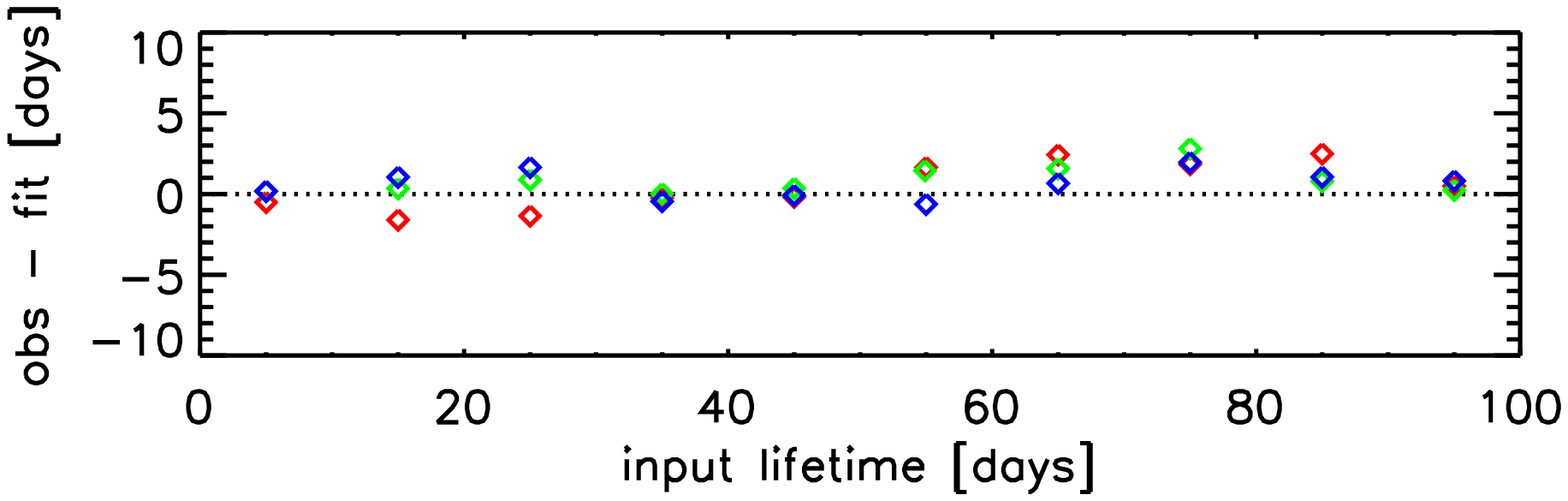}
\end{minipage}
\caption{Top: same as Fig.~\ref{lifeavgobs} for the `merge' case with height-to-background ratio ($H/B$) of the order of 10, 25 and 50 indicated in red, green and blue, respectively. The black solid line indicates the polynomial used to obtain the `unbiased' lifetime. Bottom: the difference between the `observed' lifetimes and the fit.}
\label{lifesn}
\end{figure}

The results immediately showed that the frequency at which the oscillation mode occurred did not have any influence on the fitting and the resulting lifetime. Therefore we computed the lifetime of each realisation at one frequency (35 $\mu$Hz) and computed the mean of these results. The standard error on the mean of the sample is computed as $\sigma / \sqrt{N}$ with $\sigma$ the standard deviation of the sample and $N$ the number of measured lifetimes over which the mean is computed. 
These are shown in Fig.~\ref{lifeavgobs}. This figure clearly shows that the measured lifetimes are equal to or lower than the lifetimes that were used for the simulations and that the offset from the one-to-one relations is larger for the `gap' and `merge' cases than for the `full' case. For the `full' timeseries the results become less reliable above $\sim$ 60 days while the `gap' and `merge' results are consistent only for lifetimes below 20 days.

The limited timespan of the observations compared to the mode lifetimes could be the cause of the discrepancy between our results and the input values. To investigate this further, we repeated the simulations and analysis for timeseries with half (`short') and twice (`long') the timespan of the initial simulations, i.e., 197 and 788 days and without a gap. We compared the resulting average lifetimes computed from the analysis of individual realisations. These are shown in Fig.~\ref{lifeavgtest1}. Indeed we see that for the `long' timeseries the results are consistent with or overestimating the input for all mode lifetimes investigated here, while for the `short' timeseries the discrepancy between input and observed lifetime start to emerge at lifetimes of about 25 days. This confirms that the total time span of the observations  ($T$) influences the measured mode lifetime ($\tau$), and that we can only measure `unbiased' mode lifetimes when (roughly) $T/\tau>10$.
We note that the `gap', the `merge' and the `short' timeseries, which have similar effective timespans, all show similar results. 

The results for the `gap' and `merge' timeseries are approximately equal for lifetimes up to $\sim$ 35 days. For higher input lifetimes the mean of the observed lifetimes for the `merge' timeseries are systematically lower than for the `gap' timeseries. The reason for this increasing difference is probably due to the different frequency resolutions and might therefore vary as a function of the (fractional) length of the gap. 

To test the influence of the length of the gap on the resulting lifetimes, we repeated the analyses as described above for timeseries with a total length of 394 days, and different gaps of 1.5, 0.5, 0.25 and 0.125 times the length of the gap in the observed data. The results are shown in Fig.~\ref{lifeavgtest2}. From these simulations it becomes clear that the deviations of the results shown for the  `gap' and `merge' timeseries increase with increasing length of the gap, i.e. they depend on the effective timespan of the data. We see that for timeseries with a shorter effective timespan we can measure only shorter `unbiased' mode lifetimes, which is in agreement with the results from the `short' and `long' timeseries.  Furthermore, the length of the gap in the observed timeseries (205 days) is longer than the lifetimes considered here. This changes when we investigate the same lifetimes for data with shorter gaps. Nevertheless, we find smooth trends between the observed and input lifetimes for all gaps and mode lifetimes considered here.

We also investigated the influence of the height-to-background ratio of the oscillation modes on the measurement of the mode lifetime. We did this with simulations of `merge' timeseries, similar to the observed timeseries we analysed, on which we added noise such that the height-to-background ratio ($H/B$) of the oscillation mode is of the order of 10, 25 and 50. In these simulations the noise has been added to the simulated timeseries, while the height of the oscillation mode ($H$) was derived from the root-mean-square of the flux ($A$) and the input mode lifetime ($\tau$) as $H=A^2\tau$. Results for these simulations with different $H/B$ are shown in Fig.~\ref{lifesn}. These results showed that the results are to a large extent insensitive to $H/B$, although the inclusion of noise in the simulations provided different results compared to results from noise-free simulations (see Fig.~\ref{lifeavgobs}). These differences are most likely due to the noise and due to the addition of an extra free parameter (background level) in the fitting.

From these simulations we conclude that we need to apply a correction to the `observed' mode lifetime to obtain the `unbiased' lifetime whether we leave the gap in the data or we merge the timeseries of the initial and long run. To avoid selection of frequencies which have high power due to the more severe sidelobes in the window function of the data with a gap, we have performed the analysis on observed power spectra of merged timeseries. The lifetimes of the observed modes then need to be corrected. For this correction we use a third order polynomial fit to the lifetimes measured from the simulations including noise:
\begin{equation}
y=5.4+0.96x-0.014x^2+7.1\cdot10^{-5}x^3,
\label{corcof}
\end{equation}
with $y$ the observed lifetime and $x$ the input lifetime as shown in Fig.~\ref{lifesn}. This shows that the fit is not exact, but that the deviation from the `observed' lifetimes is less than five days for all lifetimes and $H/B$. 
In the analysis of the observed stars we will use this correction as a look-up table to compute `unbiased' lifetimes for observed lifetimes between 5.4 and $\sim$ 35 days, i.e. the range of observed lifetimes over which the correction is computed. For any observed lifetimes outside this range we can infer only lower or upper limits.

\section{Individual stars}
From the red-giant candidates selected from near-IR colours and power excess as described in Sect. 2 we have four candidates for which we can fit the individual oscillations. The fitting is performed to the non-oversampled power spectra in a similar way as for the simulations (Sect. 3).  The fitting is performed in two steps. First the background, for which we use the power law fit to account for effects such as activity and granulation as described in Sect. 2 is fitted. Then, we compute a global maximum likelihood fit to all oscillation modes. In this fit the previously determined background is kept fixed, while the frequency, height and line width of each oscillation mode and the noise level are free parameters. No dependence of parameters on the degree of the mode and thus on the azimuthal order is included in the fitting. This is both because we do not know the degree of the modes and because we assume that the stars are slow rotators and the rotational splitting is negligible.

The selection of the oscillation modes is based on a statistical test of the binned power spectrum. For this test we bin the power spectrum over intervals of three frequency bins. Then we compute the probability of the power in the binned power spectrum to be due to noise. This probability is computed using a $\chi^2$ distribution in which we take the width and number of the bins into account in the degrees of freedom. Frequencies at which the probability of the power not being due to noise is larger than 80\% are selected as candidate oscillation frequencies. After performing the fit, we also verify that in the ratio of the observed to the fitted spectra no prominent peaks are left. Therefore we compute the relative height ($s$) for the investigated frequency range for which the probability of observing at least one spike with this height due to noise is 10\%, following the formulation by \citet{chaplin2002} and references therein.

From the resulting height ($H$) and width ($\Gamma$) for each oscillation frequency we also compute the root-mean-squared amplitude ($A=\sqrt{\pi H \Gamma}$). This parameter is a measure of the total power of a mode and reflects the balance between the damping and the excitation of the mode. $H$ and $\Gamma$ are not independent and this is taken into account in the computation of the error in $A$ using a correlation coefficient of $-$0.9 following the method described by \citet{chaplin2000}. 

\subsection{CoRoT 102732890}
\begin{figure*}
\begin{minipage}{\linewidth}
\centering
\includegraphics[width=\linewidth]{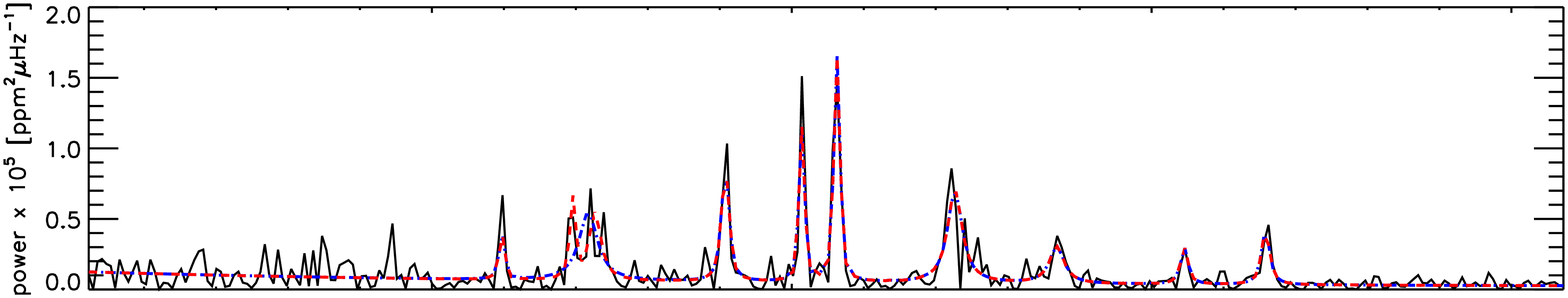}
\end{minipage}
\hfill
\begin{minipage}{\linewidth}
\centering
\includegraphics[width=\linewidth]{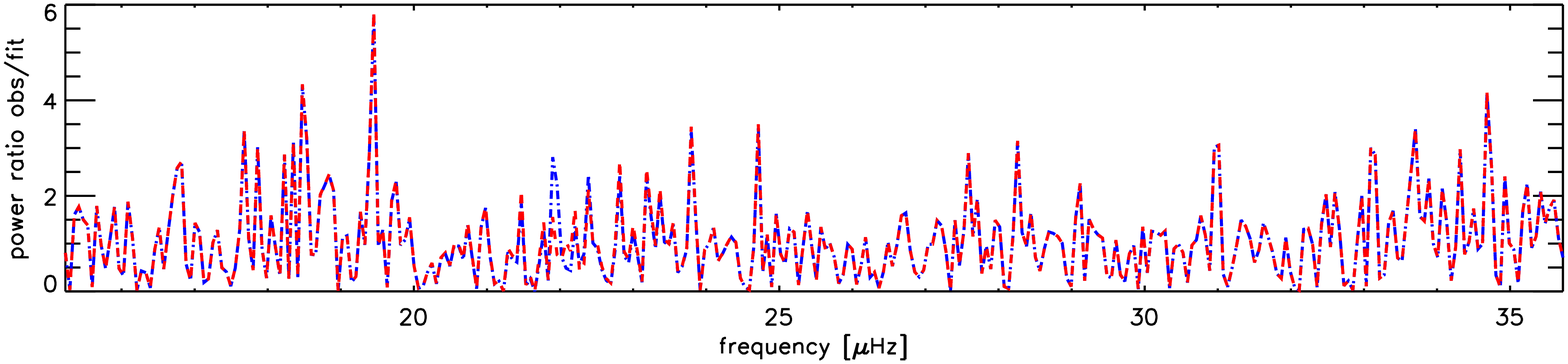}
\end{minipage}
\caption{Top: power spectrum for CoRoT 102732890 (black solid line), with a fit for nine oscillation modes (blue dash-dot line) and a fit with ten oscillation modes (red dashed line). Bottom: ratio between the observed power spectrum and the fit with nine frequencies and ten frequencies (same colours and linestyles as in the top panel). The value of $s$ for this frequency interval is 8.1. See text in Section 4 for a description of $s$.}
\label{32890_10}
\end{figure*}

CoRoT 102732890 is a star with frequencies showing maximum oscillation power at about 26 $\mu$Hz. Our statistical test indicated nine frequency intervals to have more than 80\% probability to be due to signal.  We fit for these frequencies and the results are shown in Fig.~\ref{32890_10} together with the ratio of the observed power spectrum to the fit. Indeed, in the fitted region no clear oscillation modes are present anymore. The MERIT$_{\rm red}$ is also below one (0.79), which indicates a good fit. However, one could argue that the power around 22 $\mu$Hz is not due to one oscillation mode but due to two modes. Therefore, we also perform a fit with ten frequencies. According to the MERIT$_{\rm red}$ this is also a good fit (0.79). The maximum likelihood is (slightly) higher for the fit with ten frequencies compared to the fit with nine frequencies.  We perform a likelihood ratio test \citep[e.g.,][]{appourchaux1995} to see if the increase in the maximum likelihood justifies the addition of three free parameters. 
In this particular case we see that there is a 35\% probability to get this increase without the addition of an oscillation mode. 
From this probability we cannot draw any firm conclusions and therefore we continue the analysis using both nine and ten frequencies.
The values used for the fits are listed in Table~\ref{32890_tab} together with the computed amplitude and `unbiased' mode lifetime. 

\begin{table*}
\begin{minipage}{\linewidth}
\caption{Parameters for the fit to CoRoT 102732890. The result between '()' is the oscillation mode at $\sim$ 22 $\mu$Hz in the fit with nine frequencies, while the results for the interpretation of two oscillation modes at $\sim$ 22 $\mu$Hz are indicated in square brackets. The frequencies, linewidth and height are fitted parameters, the height-to-background ratio ($H/B$) is measured as the observed height of the oscillation mode in the power spectrum over the fitted background. The amplitudes and `unbiased' lifetimes are derived from the fitted parameters as described in the text.}
\label{32890_tab}
\centering
\begin{tabular}{cccc|cc}
\hline\hline
frequency & linewidth  & height  &  $H/B$  & amplitude & lifetime \\
 $\mu$Hz  &  $\mu$Hz & 10$^5$ ppm$^2$$\mu$Hz$^{-1}$ &  & ppm & days\\
\hline
20.96 $\pm$ 0.033 & 0.026 $\pm$ 0.003 & 0.5 $\pm$ 0.1 & 9.7 & 64 $\pm$ 26 & $>$100\\
(22.2 $\pm$ 0.2 & 0.27 $\pm$ 0.04 & 0.20 $\pm$ 0.02 & 11.6 & 130 $\pm$ 33 & 10.0$^{+3.6}_{-2.4}$)\\
$[$22.0 $\pm$ 0.1 & 0.040 $\pm$ 0.004 & 0.37 $\pm$ 0.05 & 11.4 & 69 $\pm$ 17 & $>$100$]$\\
$[$22.3 $\pm$ 0.1 & 0.16 $\pm$ 0.03 & 0.22 $\pm$ 0.04 & 11.7 & 103 $\pm$ 30 & 28.3$^{+30.0}_{-8.7}$$]$\\
24.08 $\pm$ 0.06 & 0.07  $\pm$ 0.01 & 0.6 $\pm$ 0.1 & 19.5 & 107 $\pm$ 38 & $>$100\\
25.14 $\pm$ 0.03 & 0.036 $\pm$ 0.003  & 0.7 $\pm$ 0.1 & 30.9 & 87 $\pm$ 27 & $>$100\\
25.62 $\pm$ 0.04 & 0.050 $\pm$ 0.002 & 0.93 $\pm$ 0.05 & 30.8 & 120 $\pm$ 20 & $>$100\\
27.27 $\pm$ 0.07 & 0.21 $\pm$ 0.01 & 0.3 $\pm$ 0.1 & 20.3 & 136 $\pm$ 83 & 16.1$^{+1.6}_{-1.4}$\\
28.7 $\pm$ 0.6 & 0.2 $\pm$ 0.1 & 0.11 $\pm$ 0.05 & 9.9 & 85 $\pm$ 37 & 17.7$^{+82.3}_{-9.6}$\\
30.45 $\pm$ 0.07 & 0.079 $\pm$ 0.005 & 0.14 $\pm$ 0.01 & 8.4 & 59 $\pm$ 11 & $>$100\\
31.6 $\pm$ 0.1 & 0.084 $\pm$ 0.008 & 0.22 $\pm$ 0.03 & 14.1 & 76 $\pm$ 18 & $>$100\\
\hline
\end{tabular}
\end{minipage}
\end{table*}

\begin{figure}
\begin{minipage}{\linewidth}
\centering
\includegraphics[width=\linewidth]{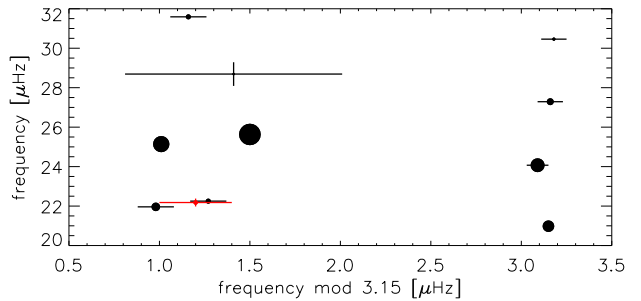}
\end{minipage}
\caption{Echelle diagram of CoRoT 102732890, using ten frequencies in black symbols, with a size proportional to the height of the fitted oscillations. When the two modes at $\sim$ 22 $\mu$Hz are replaced by one mode then the two lower left black symbols are replaced by the red symbol.}
\label{32890_echelle}
\end{figure}

From the observed $\nu_{\rm max}$ and Eq.~\ref{dnunumax}, we expect a large separation $\Delta\nu$ of about 3.2~$\mu$Hz.  Indeed when we plot the ten frequencies in an \'{e}chelle diagram with a folding frequency of 3.15 $\mu$Hz, the different mode sequences became visible (see Fig.~\ref{32890_echelle}). 

Inspection of Fig.~\ref{32890_echelle} shows that the modes are located in two roughly vertical regions. The interpretation of this is not unambiguous. One could argue that the broad ridge consists of two ridges, namely the $\ell$~=~2 and $\ell$~=~0 ridge, which seems to be the case for the fit with ten frequencies (black symbols in Fig.~\ref{32890_echelle}). We could also interpret the broad ridge as the $\ell$~=~1 ridge where we see more scatter due to less efficient trapping of the modes, (see e.g., \citet{bedding2010}, and \citet{dupret2009} for more details on mode trapping). The latter might be the case for the fit with nine frequencies as the two bottom black symbols of the broad ridge are then replaced by the red point.

\subsection{CoRoT 102788308}
\begin{figure*}
\begin{minipage}{\linewidth}
\centering
\includegraphics[width=\linewidth]{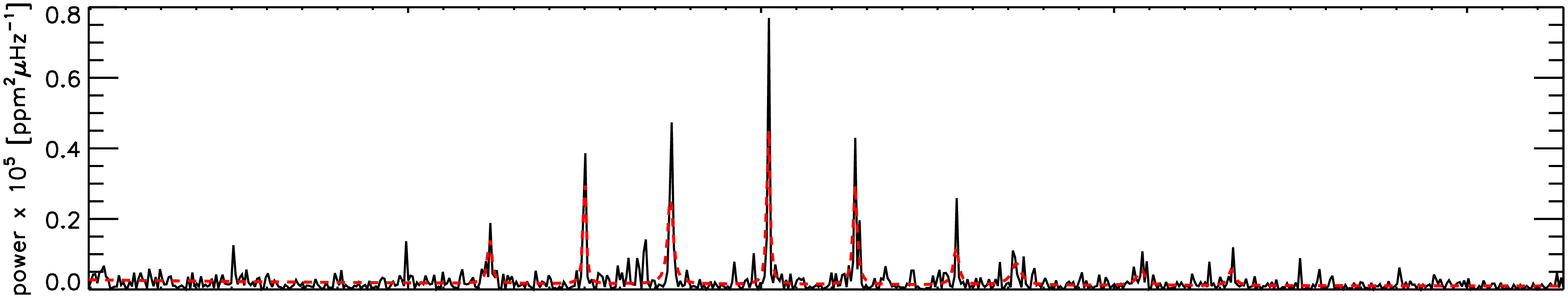}
\end{minipage}
\hfill
\begin{minipage}{\linewidth}
\centering
\includegraphics[width=\linewidth]{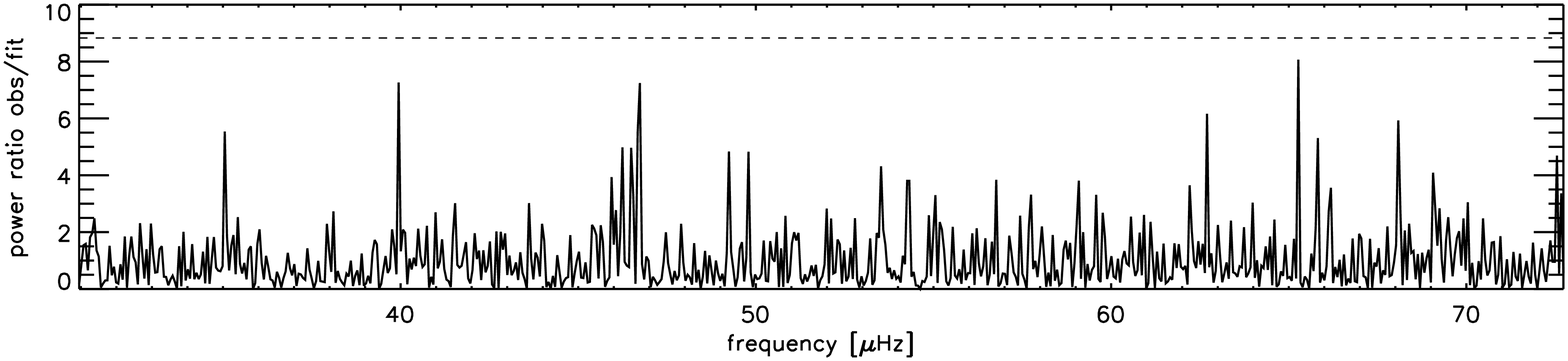}
\end{minipage}
\caption{Top: power spectrum for CoRoT 102788308 (black solid line), with a fit for nine oscillation modes (red dashed line). Bottom: ratio between the observed power spectrum and the fit. The dashed horizontal line indicates $s$ (see text for more details).}
\label{88308_new}
\end{figure*}

\begin{figure}
\begin{minipage}{\linewidth}
\centering
\includegraphics[width=\linewidth]{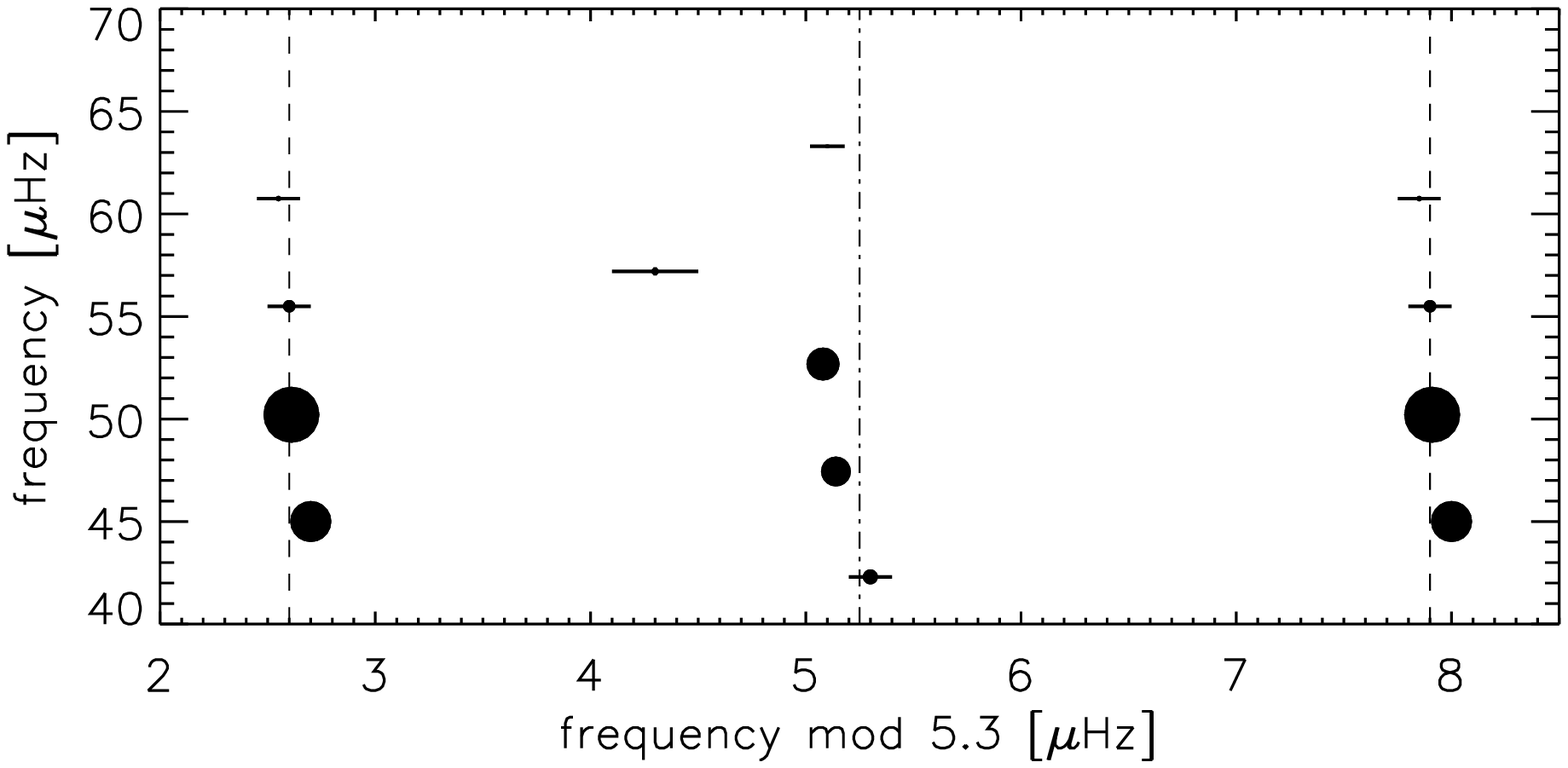}
\end{minipage}
\caption{Echelle diagram of CoRoT 102788308, with symbol sizes proportional to the height of the fitted oscillations. The dashed lines indicate the same ridge with the dashed-dotted line the midpoint between them.}
\label{88308_echelle}
\end{figure}

CoRoT 102788308 has a strikingly clear and regular pattern around a frequency of about 50 $\mu$Hz. As seen in Fig.~\ref{88308_new} and \ref{88308_echelle} for the power spectrum and \'{e}chelle diagram, respectively, and Table~\ref{88308_tab} for the fitted parameters. The MERIT$_{\rm red}$ of the global fit to the power spectrum is 1.2, which indicates that this is a reasonable fit to the data. When looking at the ratio of the observations to the fit we indeed still see some structure, with a main feature at about 46.3 $\mu$Hz. At this frequency we see for many CoRoT stars the fourth harmonic of a one day feature. Therefore we do not fit this structure in this case, although we cannot exclude that part of this signal is due to stellar oscillations. In addition, the peaks at frequencies about 35, 40, 65 and 68 $\mu$Hz might be due to oscillations, but the height to background ratio is not large enough to confirm that and therefore these are not taken into account in the fit.

\begin{table*}
\begin{minipage}{\linewidth}
\caption{Parameters for the fit to CoRoT 102788308. The frequencies, linewidth and height are fitted parameters, the height-to-background ratio ($H/B$) is measured as the observed height of the oscillation mode in the power spectrum over the fitted background. The amplitudes and `unbiased' lifetimes are derived from the fitted parameters as described in the text.}
\label{88308_tab}
\centering
\begin{tabular}{cccc|cc}
\hline\hline
frequency & linewidth  & height  & $H/B$ & amplitude & lifetime\\
 $\mu$Hz  &  $\mu$Hz & 10$^5$ ppm$^2$$\mu$Hz$^{-1}$ & & ppm & days\\
\hline
42.3 $\pm$ 0.1 & 0.15 $\pm$ 0.02 & 0.048 $\pm$ 0.007 & 11.0 & 46 $\pm$ 13 & 33.2$^{+25.1}_{-8.6}$\\
45.00 $\pm$ 0.06 & 0.083 $\pm$ 0.004 & 0.132 $\pm$ 0.009 & 24.2 & 59 $\pm$ 10 & $>$100\\
47.43 $\pm$ 0.06 & 0.175 $\pm$ 0.009 & 0.096 $\pm$ 0.006 & 32.0 & 73 $\pm$ 13 & 23.2$^{+2.8}_{-2.3}$\\
50.21 $\pm$ 0.04 & 0.097 $\pm$ 0.005 & 0.18 $\pm$ 0.01 & 55.6 & 74 $\pm$ 13 & $>$100\\
52.66 $\pm$ 0.06 & 0.137 $\pm$ 0.007 & 0.106 $\pm$ 0.007 & 32.7 & 67 $\pm$ 12 & 44.4$^{+13.9}_{-7.0}$\\
55.5 $\pm$ 0.1 & 0.17 $\pm$ 0.01 & 0.040 $\pm$ 0.004 & 20.9 & 47 $\pm$ 10 & 24.7$^{+3.6}_{-2.8}$\\
57.2 $\pm$ 0.2 & 0.3 $\pm$ 0.2 & 0.023 $\pm$ 0.002 & 9.1 & 49 $\pm$  36 & 8.1$^{+91.9}_{-6.0}$\\
60.8 $\pm$ 0.1 & 0.27 $\pm$ 0.02 & 0.017 $\pm$ 0.001 & 9.5 & 38 $\pm$ 8 & 10.0$^{+1.6}_{-1.3}$\\
63.33 $\pm$ 0.08 & 0.21 $\pm$ 0.04 & 0.01 $\pm$ 0.01 &10.9 & 33 $\pm$ 23 & 16.1$^{+8.5}_{-4.6}$\\
\hline
\end{tabular}
\end{minipage}
\end{table*}

From the $\nu_{\rm max}$ of about 50 $\mu$Hz and Eq.~\ref{dnunumax}, we expect a $\Delta \nu$ of about 5.4 $\mu$Hz. Indeed we find $\Delta \nu$ = 5.3 $\mu$Hz by folding the frequencies. The \'{e}chelle diagram shows two very clear ridges, i.e., the $\ell$ = 0 and $\ell$ = 1 ridges (although the actual degree of the ridges is not identified) with one point outside the central ridge in Fig.~\ref{88308_echelle}. This mode has most likely been shifted due to an avoided crossing and in that case the centre ridge is the $\ell$~=~1 ridge. 

\subsection{CoRoT 102762620}

\begin{figure*}
\begin{minipage}{\linewidth}
\centering
\includegraphics[width=\linewidth]{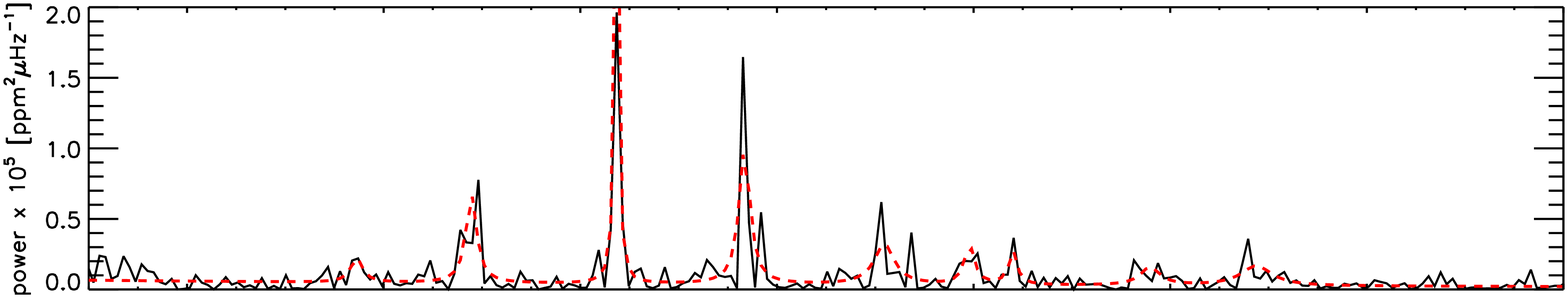}
\end{minipage}
\hfill
\begin{minipage}{\linewidth}
\centering
\includegraphics[width=\linewidth]{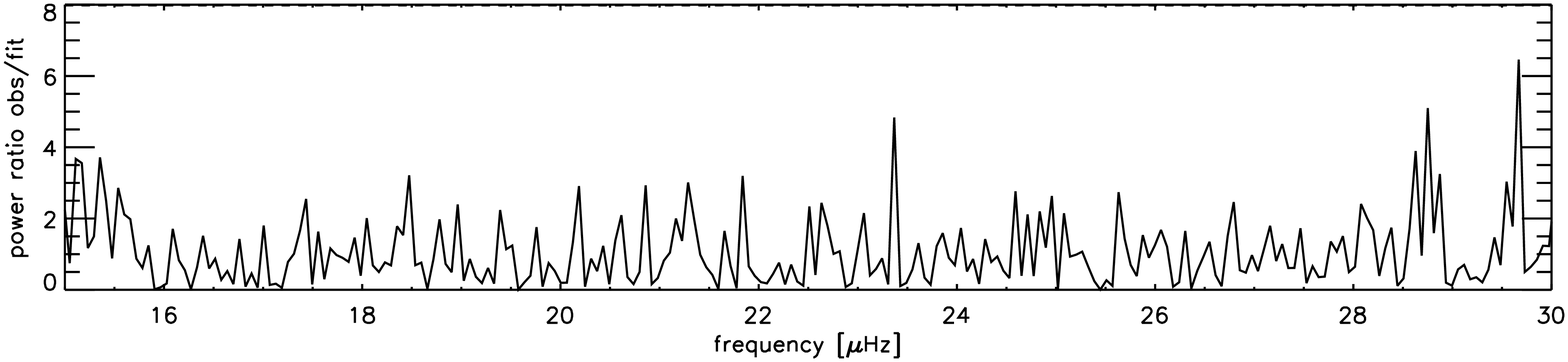}
\end{minipage}
\caption{Top: power spectrum for CoRoT 102762620 (black solid line), with a fit for nine oscillation modes (red dashed line). Bottom: ratio between the observed power spectrum and the fit. The value of $s$ for this frequency interval is 8.0 (horizontal dashed line).}
\label{62620_new}
\end{figure*}

\begin{table*}
\begin{minipage}{\linewidth}
\caption{Same as Table~\ref{88308_tab} but for parameters of the fit to CoRoT 102762620.}
\label{62620_tab}
\centering
\begin{tabular}{cccc|cc}
\hline\hline
frequency & linewidth  & height  & $H/B$ & amplitude & lifetime\\
 $\mu$Hz  &  $\mu$Hz & 10$^5$ ppm$^2$$\mu$Hz$^{-1}$ & & ppm & days\\
\hline
17.7 $\pm$ 0.2 & 0.14 $\pm$ 0.03 & 0.7 $\pm$ 0.1 & 4.5 & 54 $\pm$ 17 & 40.9$^{+57.7}_{-16.3}$ \\
18.9 $\pm$ 0.2 & 0.13 $\pm$ 0.01 & 0.24 $\pm$ 0.03 & 17.6 & 98 $\pm$ 24 & 58.3$^{+27.7}_{-17.4}$\\
20.37 $\pm$ 0.07 & 0.041 $\pm$ 0.004 & 1.4 $\pm$ 0.1 & 50.6 & 136 $\pm$ 30 & $>$100\\
21.7 $\pm$ 0.2 & 0.13 $\pm$ 0.01 & 0.37 $\pm$ 0.03 & 47.2 & 124 $\pm$ 26 & 58.3$^{+27.7}_{-17.4}$\\
23.1 $\pm$ 0.1 & 0.3 $\pm$ 0.3 & 0.10 $\pm$ 0.06 & 19.8 & 89 $\pm$ 77 & 8.1$^{+ 91.9}_{-7.2}$\\
24.0 $\pm$ 0.1 & 0.13 $\pm$ 0.06 & 0.10 $\pm$ 0.03 & 12.5 & 63 $\pm$ 27 & 58.3$^{+ 41.7}_{-38.7}$\\
24.40 $\pm$ 0.07 & 0.07 $\pm$ 0.01 & 0.13 $\pm$ 0.02 & 12.8 & 53 $\pm$ 13 & $>$100\\
25.8 $\pm$ 0.6 & 0.241 $\pm$ 0.005 & 0.047 $\pm$ 0.006 & 8.0 & 59 $\pm$ 21 & 12.4$^{+0.5}_{-0.5}$\\
26.9 $\pm$ 0.3 & 0.38 $\pm$ 0.06 & 0.05 $\pm$ 0.03 & 14.8 & 79 $\pm$ 57 & 4.8$^{+2.3}_{-1.6}$\\
\hline
\end{tabular}
\end{minipage}
\end{table*}

\begin{figure}
\begin{minipage}{\linewidth}
\centering
\includegraphics[width=\linewidth]{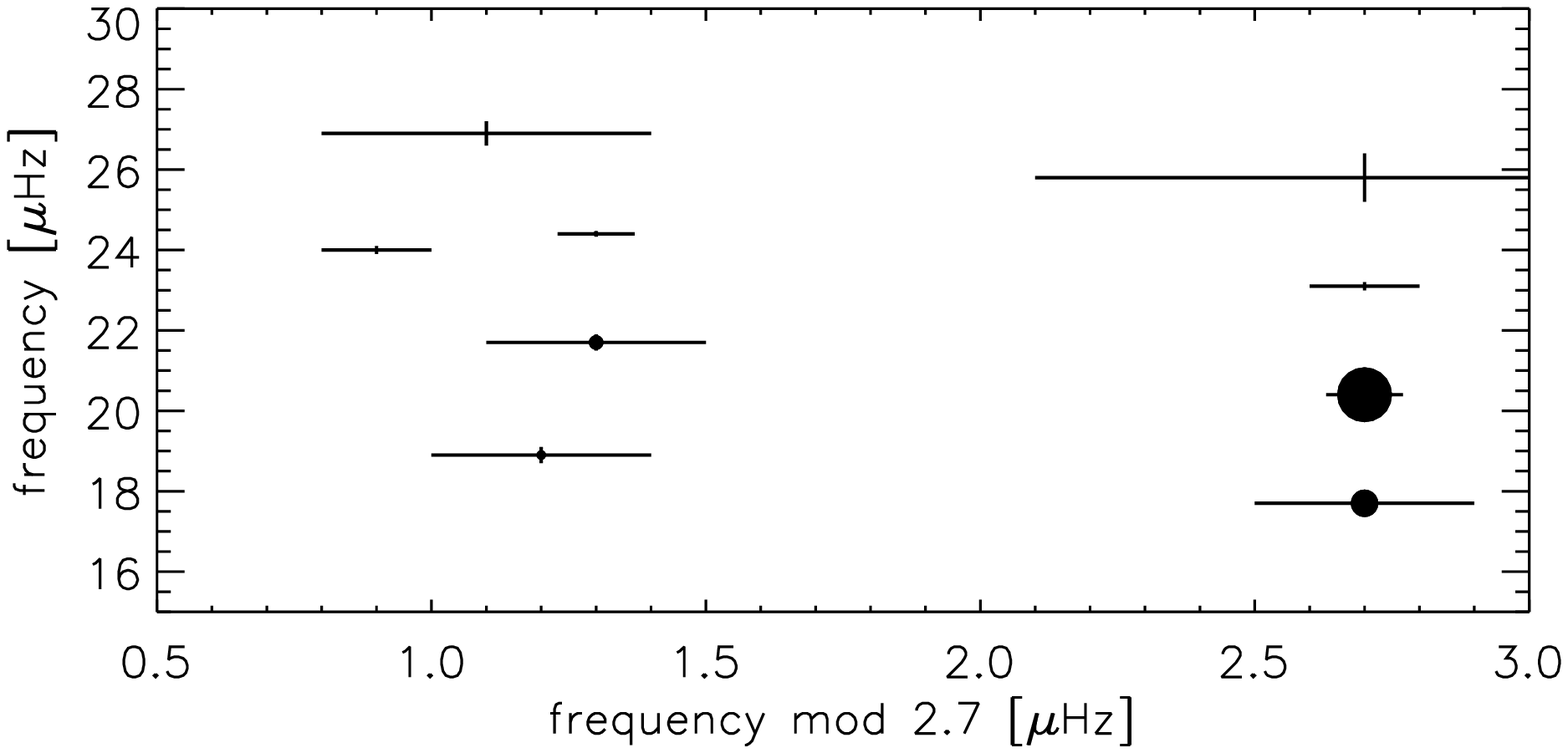}
\end{minipage}
\caption{Echelle diagram of CoRoT 102762620, with symbol sizes proportional to the height of the fitted oscillations.}
\label{62620_echelle}
\end{figure}

CoRoT 102762620 shows oscillation frequencies at slightly lower frequencies than CoRoT 102732890. With the statistical test of the smoothed power spectrum we find oscillations roughly between 17 and 27 $\mu$Hz (see Fig.~\ref{62620_new} and Table~\ref{62620_tab} for the fitting parameters).  

From the frequency of maximum oscillation power of about 22 $\mu$Hz we predict a $\Delta \nu$ of about 2.8 $\mu$Hz. From the observations we find a consistent value of 2.7 $\mu$Hz for this star. This results in one narrow ridge and a second ridge which consist either of two ridges relatively close together or one ridge with larger scatter.

\subsection{CoRoT 102767771}

\begin{figure*}
\begin{minipage}{\linewidth}
\centering
\includegraphics[width=\linewidth]{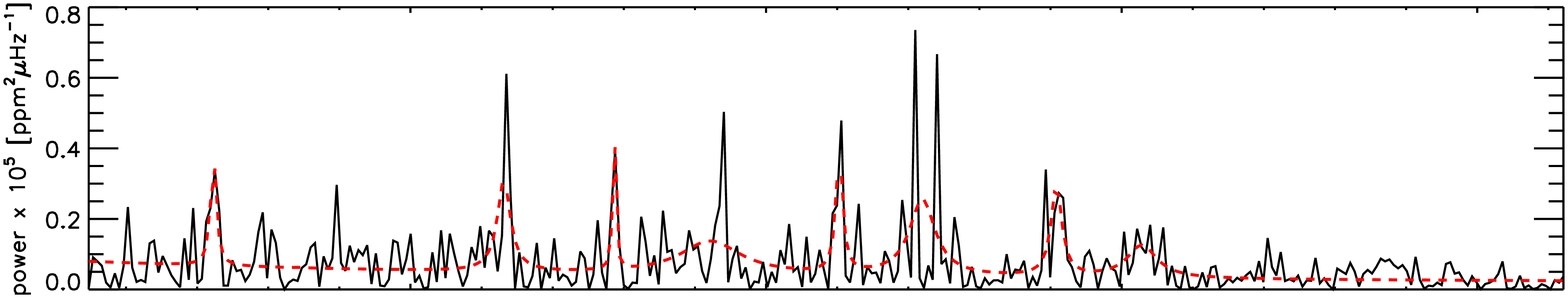}
\end{minipage}
\hfill
\begin{minipage}{\linewidth}
\centering
\includegraphics[width=\linewidth]{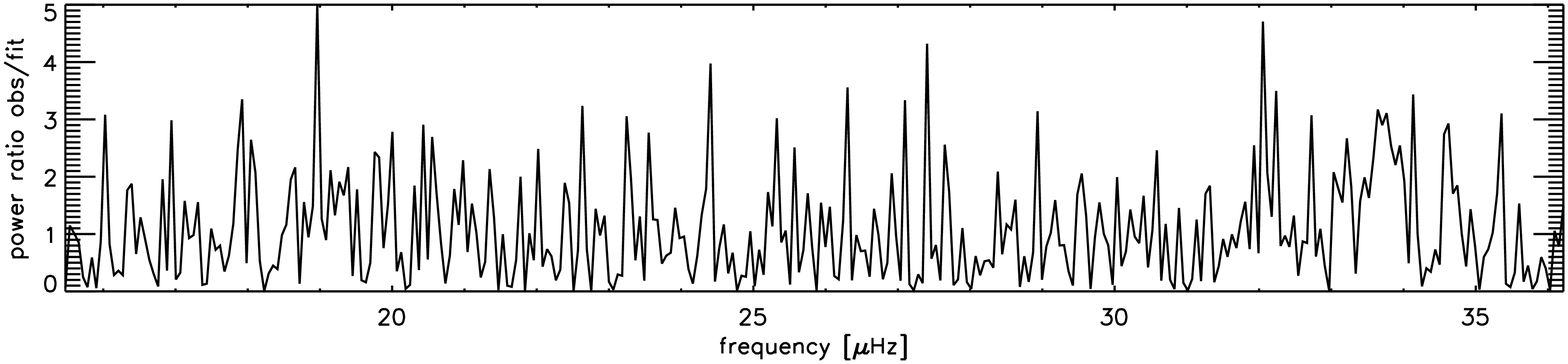}
\end{minipage}
\caption{Top: power spectrum for CoRoT 102767771 (black solid line), with a fit for eight oscillation modes (red dashed line). Bottom: ratio between the observed power spectrum and the fit. The value of $s$ for this frequency interval is 8.1.}
\label{67771_new}
\end{figure*}

\begin{table*}
\begin{minipage}{\linewidth}
\caption{Same as Table~\ref{88308_tab} but for parameters of the fit to CoRoT 102767771.}
\label{67771_tab}
\centering
\begin{tabular}{cccc|cc}
\hline\hline
frequency & linewidth  & height  & $H/B$ & amplitude & lifetime\\
 $\mu$Hz  &  $\mu$Hz & 10$^5$ ppm$^2$$\mu$Hz$^{-1}$ & & ppm & days\\
\hline
17.23 $\pm$ 0.09 & 0.088 $\pm$ 0.02 & 0.12 $\pm$ 0.05 & 5.1 & 59 $\pm$ 25 & $>$100\\
21.31 $\pm$ 0.09 & 0.22 $\pm$ 0.02 & 0.10 $\pm$ 0.01 & 12.7 & 82 $\pm$ 24 & 14.8$^{+3.0}_{-2.3}$\\
22.86 $\pm$ 0.09 & 0.04 $\pm$ 0.01 & 0.2 $\pm$ 0.1 & 9.1 & 50 $\pm$ 24 & $>$100\\
24.21 $\pm$ 0.09 & 1.1 $\pm$ 0.2 & 0.03 $\pm$ 0.01 & 12.6 & 108 $\pm$ 40 & $<$5\\
26.0 $\pm$ 0.1 & 0.11 $\pm$ 0.01 & 0.13 $\pm$ 0.02 & 13.2 & 66 $\pm$ 19 & 98.6$^{>100}_{-12.6}$\\
27.2 $\pm$ 0.3 & 0.44 $\pm$ 0.06 & 0.08 $\pm$ 0.01 & 21.5 & 105 $\pm$ 26 & 3.3$^{+1.6}_{-1.1}$\\
29.1 $\pm$ 0.2 & 0.18 $\pm$ 0.03 & 0.10 $\pm$ 0.02 & 10.9 & 74 $\pm$ 23 & 21.9$^{+11.4}_{-5.7}$\\
30.3 $\pm$ 0.3 & 0.51 $\pm$ 0.07 & 0.034 $\pm$ 0.008 & 6.1 & 74 $\pm$ 26 & 2.0$^{+1.3}_{-1.0}$\\
\hline
\end{tabular}
\end{minipage}
\end{table*}

\begin{figure}
\begin{minipage}{\linewidth}
\centering
\includegraphics[width=\linewidth]{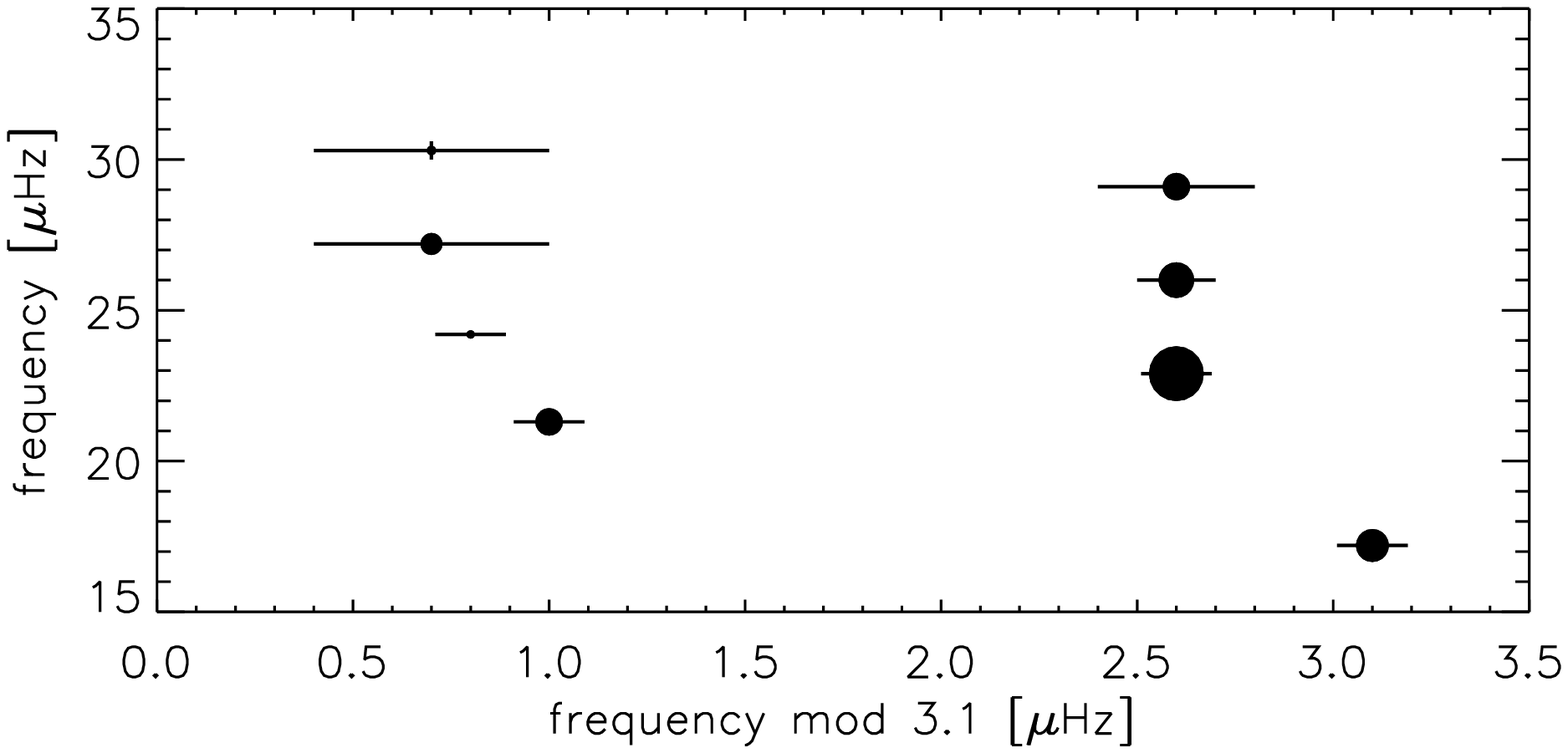}
\end{minipage}
\caption{Echelle diagram of CoRoT 102767771, with symbol sizes proportional to the height of the fitted oscillations.}
\label{67771_echelle}
\end{figure}

CoRoT 102767771 shows oscillations at about the same frequencies as 102732890, although the power of the oscillations in this star is less than for 102732890. We find a similar value for $\Delta \nu$ of 3.1 $\mu$Hz (Fig.~\ref{67771_echelle}) as for 102732890, which is in line with the predictions.  The \'{e}chelle diagram shows two clear ridges and only the mode with lowest frequency seems to be located outside the right ridge. From the current analysis we cannot say whether this ridge is curved and this mode is located on the ridge, whether the frequency is altered due to avoided crossing, or whether this is induced by noise. 

\section{Discussion and conclusions}

\subsection{Simulations}
Simulations of long (a few hundred days) timeseries of data, with different timespans and different length of gaps reveal that for short mode lifetimes we can recover the input lifetime of the mode, irrespective of the presence of a gap in the data, or whether we merge the data. For longer lifetimes we see a general under estimation of the `observed' mode lifetimes compared to the input lifetimes, in which the `observed' mode lifetime seems to reach a maximum value.  The investigations of longer and shorter timeseries reveal that for roughly $\tau < 0.1T$ the observed lifetimes are consistent with the `unbiased' lifetime. For longer lifetimes a correction has to be applied, although the uncertainty in the resulting values increases with increasing lifetimes.
We also investigated the trends in the observed lifetimes as a function of the length of the gap and whether merging the data has a large influence. As expected the shorter the gap the better. Furthermore, the results for the data with a gap and merged together differ for longer lifetimes, but as a correction is necessary in both cases, we concluded that we could merge the data of the initial run and second long run of CoRoT observations. This has the advantage that sidelobes of the window function are avoided


\begin{figure}
\begin{minipage}{\linewidth}
\centering
\includegraphics[width=\linewidth]{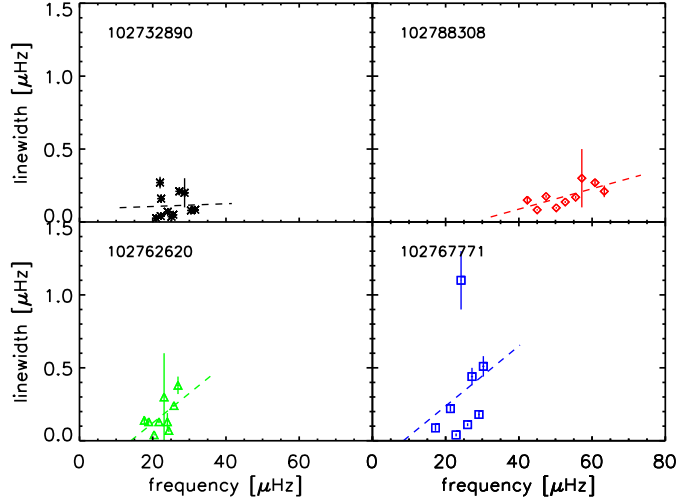}
\end{minipage}
\caption{Linewidth of the oscillation modes of the four stars studied in detail as a function of frequency. The dashed lines indicated linear fits to the results.}
\label{ltimefreq}
\end{figure}

\begin{figure}
\begin{minipage}{\linewidth}
\centering
\includegraphics[width=\linewidth]{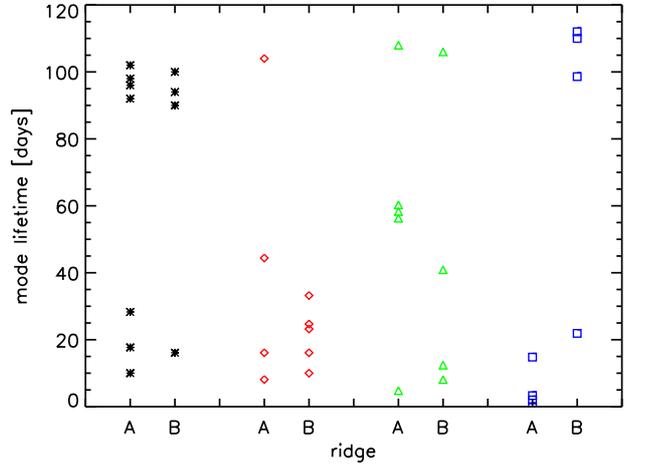}
\end{minipage}
\caption{Mode lifetimes for ridge `A' and `B'  in the \'echelle diagrams of the four stars separately. The degree of the ridges are not known and the assignment of `A' and `B' to a ridge is arbitrary. The symbols are the same as in Fig.~\ref{ltimefreq}. The oscillation modes for which we could only find a lower limit are plotted with a lifetime around 100 days, but with a small offset. An offset is also introduced in two of the three modes with a lifetime of 58.3 days (102762620).}
\label{ltimedegree}
\end{figure}

\subsection{Observations}
For four red-giant stars we have data from the initial and second long run of CoRoT at our disposal with high enough signal-to-noise ratio to perform a global fit to the individual oscillation frequencies. The oscillation modes were identified using a statistical test of the binned power spectrum. In this global fit the frequency, linewidth and height are free parameters for each identified oscillation mode.  From the parameters obtained from the fitting we also compute the root-mean-squared amplitude and the mode lifetime. The latter includes a correction using the third order polynomial obtained from the averaged mode lifetimes of the simulations including noise of merged data.

\subsubsection{$\ell$~=~2 modes}
Interestingly for none of the four stars did we find a clear $\ell$~=~2 ridge in the \'echelle diagram. For 102732890 and 102762620 one ridge is broad and could be interpreted as an $\ell$~=~0 and $\ell$~=~2 ridge but for 102788308 and 102767771 only two ridges, separated by $\sim$0.5$\Delta \nu,$ are present. The non-detection of the $\ell$~=~2 modes could possibly be due to a lack of frequency resolution, in which case the $\ell$~=~0 and $\ell$~=~2 modes are interpreted as one oscillation or due to the lack of height in the power spectrum. We investigate both possibilities.

\begin{table}
\begin{minipage}{\linewidth}
\caption{Global parameters for the four stars: frequency of maximum oscillation power ($\nu_{\rm max}$) and the large frequency separation ($\Delta\nu$ in $\mu$Hz, the ratio of the mean height of ridge `A' to ridge `B' for which the degrees are unknown, the effective temperature ($T_{\rm eff}$) in K and the asteroseismic mass and radius in M$_{\odot}$ and R$_{\odot}$, respectively. Errors are typically of the order of 1 and 0.05 $\mu$Hz for $\nu_{\rm max}$ and $\Delta \nu$, 100 K for $T_{\rm eff}$ and below 20 and 10 \% for $M$ and $R$, respectively.}
\label{sum}
\centering
\begin{tabular}{lcccccc}
\hline\hline
star & $\nu_{\rm max}$ & $\Delta\nu$ & $H_A/H_B$ & $T_{\rm eff}$ & $M$ & $R$\\
 & $\mu$Hz  &  $\mu$Hz & & K &M$_{\odot}$ & R$_{\odot}$\\
\hline
102732890 & 25.5 & 3.15 &1.1$\pm$ 0.1 & 4140 & 1.2 & 13.0\\
102788308 & 51.8 & 5.3 & 1.6 $\pm$ 0.1 & 4260 & 1.3 & 9.4\\
102762620 & 21.9 & 2.7 & 0.3 $\pm$ 0.03 & 4355 & 1.5 & 15.6\\
102767771 & 25.8 & 3.1 & 0.4 $\pm$ 0.1 & 4250 & 1.4 & 13.7\\
\hline
\end{tabular}
\end{minipage}
\end{table}

From the individual oscillations observed for 102788308, we can obtain the differences between consecutive frequencies with reasonable accuracy. We find here two typical differences: 2.4 and 2.8 $\mu$Hz. This means that the amount of offset of the $\ell$ =1 modes from the midpoints between the $\ell$ = 0 modes ($\delta \nu_{01}$) is of the order of 0.2 $\mu$Hz. If we interpret the centre ridge as $\ell$~=~1 ridge (Fig.~\ref{88308_echelle}) then this ridge is mostly located on the left side of the midpoint (dot-dashed vertical line in Fig.~\ref{88308_echelle}) of the $\ell$~=~0 modes (dashed vertical lines in Fig.~\ref{88308_echelle}). The position of the $\ell$~=~1 ridge on the left side of the midpoint between $\ell$~=~0 modes might indicate that this star is in an earlier evolutionary state and that it seems to be justified to use the prediction of the small separation from the asymptotic approximation (Eq.~\ref{asymptot}). When we use this approximation and the estimate of $\delta \nu_{01}$ of the observed frequencies, we find that $\delta \nu_{02}$ should be of the order 0.6 $\mu$Hz. The frequency at $\sim$57.2 $\mu$Hz lies by this amount outside the centre ridge but no other convincing features are present, although the frequency resolution of the dataset is sufficient to resolve this separation.

Also for 102767771 we find evidence for only two ridges and we seem to miss the $\ell$~=~2 ridge. In this case this might be caused by a combination of a lower signal-to-noise ratio and the low frequencies of the oscillations. At these low frequencies we expect also a smaller separation between the $\ell$~=~0 and $\ell$~=~2 modes compared to the 0.6~$\mu$Hz estimated for 102788308 and the frequency resolution of the spectrum is therefore more critical to detect them, although the frequency resolution of $\sim$0.06 $\mu$Hz should be sufficient. If the modes are very wide, i.e., short mode lifetimes or unresolved mode pairs, a single observed power excess hump might actually consist of two modes. Fitting the two modes as one will overestimate the line width considerably and the correction to compute the `unbiased' mode lifetime (Eq.~\ref{corcof}) might also no longer be valid, as this is determined for a single oscillation mode. 

The theoretical computations by \citet{dupret2009} predict that the trapping of $\ell$~=~2 modes is more efficient than the trapping of the $\ell$~=~1 modes due to the increased size of the evanescent region for modes of higher degree. Due to this more efficient trapping the $\ell$~=~2 modes are predicted to have similar heights in the power spectrum as radial modes. In this prediction the visibility effects are not taken into account. For the Sun it is known that  modes with degrees $\ell$~=~0, 1, 2 have relative heights 1.0, 1.5 and 0.5. These relative heights depend on limb darkening and might be different for red giants compared to the Sun. For the stars investigated here we computed  the mean of the fitted heights per ridge and the ratio of these mean values. The results are listed in Table~\ref{sum} and cover a considerable range (note that the inverse of the ratio should also be considered as the degree of the ridges `A' and `B' is unknown). This might indicate that the relative visibilities of modes with different degrees in red giants are different from that of the Sun and this might cause the non-detection of $\ell$~=~2 modes, although this would be in contradiction with red giants for which $\ell$~=~2 modes have been observed \citet[e.g.,][]{deridder2009,carrier2010}.

It might also be possible that the lifetimes of the $\ell$~=~2 modes are long and that these modes are not resolved. Unresolved non-radial modes with long lifetimes have smaller heights than the closest radial mode \citep{dupret2009}. With longer observing runs, these modes can be resolved, which will improve the chances to detect them.

Despite the difficulties to detect $\ell$~=~2 modes encountered for the stars investigated here, these modes have been observed in a considerable number of (but not all) red-giant stars using both CoRoT and Kepler data with shorter timespans \citep{deridder2009,carrier2010,bedding2010,huber2010,kallinger2010b}. The reason why the $\ell$~=~2 modes are not observable in the stars investigated here is not entirely clear. The fact that for some stars $\ell$~=~2 modes are observed and that for other stars, such as the ones investigated here we cannot observe these modes, might also contain information about their internal structures. The small number statistics of only four stars does not allow us to draw any firm conclusions at this stage.

\subsubsection{Mode lifetimes}
Theory predicts that mode lifetimes depend on frequency and on the degree of the mode. In general the mode lifetimes decrease for modes with increasing frequencies and are longer for modes with higher degrees \citep{dupret2009}. We test this prediction using the four stars we analysed in detail, by plotting the linewidth of each mode as a function of frequency and perform a linear fit (see Fig.~\ref{ltimefreq}). For three of the four stars there is a clear increase of the linewidth with frequency present, i.e., the mode lifetimes decrease with increasing frequency. 

Also a considerable number of lifetimes observed in the four stars are also in agreement with the $\langle \tau \rangle \sim T^{-4}_{\rm eff}$ scaling prediction presented by \citet{chaplin2009}. For the effective temperatures mentioned in Table~\ref{sum} lifetimes of the order 11-15 days can be expected. We indeed see in Fig.~\ref{ltimedegree} that many results are present in this regime.

Furthermore, we investigate the dependence of lifetimes on the degree of the modes (see Fig.~\ref{ltimedegree}). The most obvious example of this dependence is present in 102767771 where all modes in the left (`A') ridge  in the \'echelle diagram in Fig.~\ref{67771_echelle} have lifetimes shorter than 20 days, while all modes in the right (`B') ridge have lifetimes longer than 20 days. Although we could not identify the degree of the ridges `A' and `B', this result is consistent with predictions by \citet{dupret2009}, namely that modes with different degrees can have systematically different mode lifetimes. For the other stars the dependence of the mode lifetime on degree is not so clear, although the two longest mode lifetimes for 102788308 are both present in the ridge indicated with the dashed vertical line in Fig.~\ref{88308_echelle} (ridge `A'). The three modes with the same lifetime of 58 days (102762620) are also located in the same ridge. 

From the \'echelle diagrams we were not yet able to identify the degrees of the modes. Because \citet{dupret2009} predict that non-radial modes have longer lifetimes than radial modes, we use that as additional information to identify the degree of the modes. For 102767771 this would indicate that the right ridge in Fig.~\ref{67771_echelle} (ridge `B') is the $\ell$~=~1 ridge and the left ridge (`A') the $\ell$~=~0 ridge. 
 
The investigation of mode lifetimes as performed here on four CoRoT stars will benefit from longer uninterrupted timeseries on a larger number of stars. This will be possible in the near future. The Kepler satellite will be taking data of a preselected set of stars during its entire lifetime of nominally 3.5 years. This increased timespan with only short interruptions and the lower noise level of the photometric timeseries compared to CoRoT or any other facility currently in operation, will provide data which will be even better suited for investigating the mode lifetimes of solar-like oscillations in red-giant stars. Analysis of the currently available Kepler data with a timespan of $\sim$230 days is in progress.
 
 \subsubsection{Stellar parameters}
For the four stars we investigated in detail we also compute the asteroseismic mass and radius using $\Delta \nu$, $\nu_{\rm max}$ and $T_{\rm eff}$, as described by \citet{kallinger2010}. For this determination we need the effective temperatures. The effective temperatures have been derived using Exo-Dat photometric data \citep{deleuil2009} using the relations of \citet{alonso1999}, as described by \citet{baudin2010}. Because of large, systematic discrepancies between the reddening values along the various lines of sight estimated from the extinction maps of \citet{dobashi2005} and \citet{rowles2009} a representative value, $A_V$ = 0.6 mag, was adopted for all stars. In view of the quite uncertain reddening, and although we note that there is a good agreement between the values derived from optical and near-IR colour indices, only the temperatures based on 2MASS data are considered for the determination of the mass and radius. The values of these stellar parameters are listed in Table~\ref{sum}.
 
 \subsubsection{Comparison with models \citet{dupret2009}}
The frequency of maximum oscillation power provides an indication of the evolutionary state of the stars. The stars 102732890 and 102767771 have a $\nu_{\rm max}$ of about 26 $\mu$Hz. Stars with $\nu_{\rm max}$ roughly between 25 and 40 $\mu$Hz are commonly observed \citep{hekker2009} and are to a large extent low-mass He-burning stars in the red clump \citep{miglio2009}. Both 102732890 and 102767771 are therefore most likely to be red-clump stars. The oscillations in 102762620 are centred around 22 $\mu$Hz, which is comparable with model E of \citet{dupret2009}. This is model of a 2 M$_{\odot}$ star high in the red giant branch. For 102788308 we find oscillations with $\nu_{\rm max}$~$\sim$~50~$\mu$Hz. These frequencies are comparable to models B, D and E of \citep{dupret2009}, which represent a 2 M$_{\odot}$ star ascending the giant branch, and 3 M$_{\odot}$ H-shell and He-shell burning models respectively. The models all have slightly higher masses than the asteroseismic masses derived from the observations. The exact positions of the stars in the H-R diagram is as yet unknown, due to the difference in mass between the models and the observations, although we can still conclude that these four stars are likely to be in different evolutionary phases on the red-giant branch. 

The frequency pattern of 102788308 is strikingly regular and the linewidths and trend of the linewidth with frequency for this star are different from that of the others. This is most likely due to the different evolutionary state of this star. This will be further investigated using detailed modelling of these stars, which will be presented in subsequent publications.

\acknowledgements
SH, WJC and YE acknowledge financial support from the UK Science and Technology Facilities Council (STFC). T.M. acknowledges financial 
support from Belspo for contract PRODEX-GAIA DPAC. This research has made use of the Exo-Dat database, operated at LAM-OAMP,
Marseilles, France, on behalf of the CoRoT/Exoplanet program. We would like to thank the referee R. Gilliland for his useful report that improved the paper.

\bibliographystyle{aa}
\bibliography{corotIRLR}

\end{document}